\documentstyle[aps,prl,psfig]{revtex}
\newcommand{\req}[1]{(\ref{#1})}
\newcommand{\be}{\begin{equation}}
\newcommand{\ee}{\end{equation}}
\newcommand{\bea}{\begin{eqnarray}}
\newcommand{\eea}{\end{eqnarray}}
\newcommand{\dd}{\textrm{d}}
\newcommand{\pr}[1]{\left(#1\right)}
\newcommand{\cro}[1]{\left[#1\right]}
\newcommand{\acc}[1]{\left\{#1\right\}}
\newcommand{\avg}[1]{\langle{#1}\rangle}
\newcommand{\Avg}[1]{\Big<{#1}\Big>}
\newcommand{\eps}{\epsilon}
\newcommand{\beps}{\bar\epsilon_{th}}

\newcommand{\erf}{\textrm{erf}}
\newcommand{\sgn}{\textrm{sgn}}

\begin{document}
\title{Relevance of memory in Minority Games}
%\twocolumn[\hsize\textwidth\columnwidth\hsize\csname
%@twocolumnfalse\endcsname
\author{Damien Challet$^{(1)}$ and Matteo Marsili$^{(2)}$}
\address{ $^{(1)}$ Institut de Physique Th\'eorique, Universit\'e de Fribourg, CH-1700\\$^{(2)}$ Istituto Nazionale per la Fisica della Materia (INFM), Trieste-SISSA Unit, V. Beirut 2-4, Trieste I-34014,\\}
\date{\today}
\maketitle
%\widetext
%\textwidth       18.0 cm
%\textheight      25.0 cm
%\topmargin       -3.0 cm
%\topskip         -1.0 cm
%\oddsidemargin   -1.0 cm
%\evensidemargin  -1.0 cm                                              

\def\rit{\hbox{\it I\hskip -2pt  R}}

\newcommand{\ovl}[1]{\overline{#1}}
\newcommand{\BE}{\begin{eqnarray}}
\newcommand{\EE}{\end{eqnarray}}
\newcommand{\BEn}{\begin{eqnarray*}}
\newcommand{\EEn}{\end{eqnarray*}}
\newcommand{\barr}{\begin{array}}
\newcommand{\earr}{\end{array}}
\newcommand{\qe}{\`{e }}
\newcommand{\eq}{\'{e }}
\newcommand{\qa}{\`{a }}
\newcommand{\qo}{\`{o }}
\newcommand{\qi}{\`{\i }}
\newcommand{\bit}{\begin{itemize}}      
\newcommand{\eit}{\end{itemize}}
\newcommand{\bc}{\begin{center}}
\newcommand{\ec}{\end{center}}
\newcommand{\ben}{\begin{enumerate}}    
\newcommand{\een}{\end{enumerate}}
\newcommand{\nid}{\noindent}
\newcommand{\cl}{\centerline}
\newcommand{\nl}{\newline}
\newcommand{\ul}{\underline}

\newcommand{\de}{\partial}
\newcommand{\impl}{\Longrightarrow}
\newcommand{\To}{\longrightarrow}
\newcommand{\LRARR}{\Longleftrightarrow}

\newcommand{\Tb}{{\bf T}}
\newcommand{\Gb}{{\bf G}}
\newcommand{\Fb}{{\bf F}}
\newcommand{\xb}{{\bf x}}
\newcommand{\yb}{{\bf y}}
\newcommand{\eb}{{\bf e}}
\newcommand{\wb}{{\bf w}}
\newcommand{\ub}{{\bf u}}
\newcommand{\om}{\omega}
\newcommand{\e}{\mbox{e}}

%%%%%%%%%%%%%%%%%%%%%%%%%%%%%%%%%%%%%%%%%%%%%%%%%%%%%%%%%%%%%%%%%%%%%%%%%%%%%

\begin{abstract} 

By considering diffusion on De Bruijn graphs, we study in details the dynamics of the histories in the Minority Game, a model of competition between adaptative agents. Such graphs describe the structure of temporal evolution of $M$ bits strings, each node standing for a given string, i.e. a history in the Minority Game. We show that the frequency of visit of each history is not given by $1/2^M$ in the limit of large $M$ when the transition probabilities are biased. Consequently all quantities of the model do significantly depend on whether the histories are real, or uniformly and randomly sampled. We expose a self-consistent theory of the case of real histories, which turns out to be in very good agreement with numerical simulations. 
\end{abstract}
\bigskip

\bigskip
\section{Introduction}

The Minority Game \cite{CZ97,web}  has been designed as the most drastic possible
simplification of Arthur's El Farol's bar problem \cite{Arthur}. 
It  is believed to capture some essential and general features of
competition between adaptative agents, which is found for instance in
financial markets.  In this model, agents have to take each time step
one of two decisions; they share a common piece of information $\mu\in
\{0,\cdots,P-1\}$ that encodes the state of the world, use it to make
their choice, and those who happen to be in minority are rewarded. In
its original formulation, the piece of information is the binary
encoding of the $M$ last winning choices, hence $P=2^M$. Hence, the dynamics
of $\mu$ is coupled to the dynamics of agents.

Cavagna \cite{cavagna} claimed that all quantities of the system ``are
completely independent from the memory of the agents''. This means that
replacing the dynamics of $\mu$ induced by agents by a random
history $\mu$ drawn at random at each time step, one finds the same results. While this 
statement has turned out to be wrong for many extensions of the MG 
\cite{J99,J99.2,CMZe99,CMZe99.2}, it has been helpful as a first approximation
for the analytical understanding of the standard MG:
an exact solution for random histories has been found in
the ``thermodynamic'' limit \cite{CMZe99,CMZe99.2}.
Interestingly, this solution shows that all quantities depend on the frequencies 
$\{\rho^\mu\}$ of visit of histories. The random history case is recovered
if $\rho^\mu=1/P$, but in the real dynamics of the MG the distribution 
$\rho^\mu$ is determined by the behavior of agents (indeed modifying the
behavior of agents may have strong effects on $\rho^\mu$ as shown in ref. 
\cite{CMZe99,CMZe99.2}). 
It turns out, that the frequencies $\rho^\mu$ are 
not uniform for all parameters of the MG.

In this paper we study quantitatively this problem. The first step is
to characterize the properties of the dynamics of real histories, which amounts to
study randomly biased diffusion on De Bruijn graphs. Depending on the 
asymmetry of the bias, we quantify the deviation $\delta\rho^\mu=
\rho^\mu-1/P$ from the uniform distribution. Then we move to the MG and
quantify the bias which agents induce on the dynamics of $\mu$
in the asymmetric phase. Using a simple parameterization of
$\rho^\mu$ which is inferred from numerical data, we 
generalize the calculations of refs. \cite{CMZe99,CMZe99.2}.
This leads to a self-consistent equation between
the asymmetry of the game and the diffusion bias, which we can solve. 
The results are in excellent agreement with numerical simulations 
and show a systematic deviation from the random history MG.
Hence, our conclusion is that, even though the random history MG
is qualitatively similar to the original MG, memory is actually
not irrelevant, and one can quantify the difference between the
two cases.

\section{De Bruijn graphs}

\begin{figure}
\centerline{\psfig{file=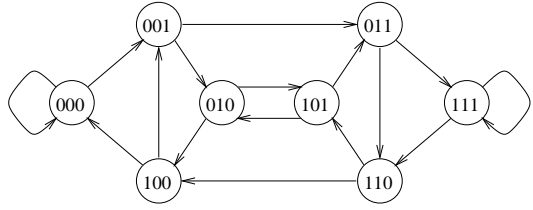,width=7cm}}
\caption{\label{debrm3}De Bruijn graph of order 3}
\end{figure}

Let us begin with the definition of some elementary concepts. A binary
sequence $\mu(t)$ of length $m$ consists of $m$ ordered elements
$\{b(t-m),\cdots,b(t-1)\}$ where $b$ is a letter belonging to the
alphabet $\{0,1\}$. $\mu(t+1)$ is obtained by adding $b(t)$ to the
right of $\mu(t)$ and erasing $b(t-m)$. Thus, for a given $\mu(t)$,
there are two possible $\mu(t+1)$, which we call ``next
neighbours''. This updating rule defines the De Bruijn graph
\cite{Bruijn} of order $m$ (see Fig. \ref{debrm3} for an example).

Let $G$ be the $P\times P$ adjacency matrix of the De Bruijn graph of
order $m$. if we adopt the convention that its elements are indiced by
the decimal value of the binary strings, that is, $\mu=0,\cdots,P-1$,
\be G_{\mu,\nu}=\delta_{[2\mu\%P],\nu}+\delta_{[2\mu\%P]+1,\nu} \ee
where $A\%B$ stands for the remainder of the division of $B$ by $A$
and $\delta_{i,j}$ is the Kronecker symbol. 
The adjacency matrix for $m=3$ is:  
\be 
G=\left(
\begin{array}{cccccccc}
1&1&0&0&0&0&0&0\\
0&0&1&1&0&0&0&0\\
0&0&0&0&1&1&0&0\\
0&0&0&0&0&0&1&1\\
1&1&0&0&0&0&0&0\\
0&0&1&1&0&0&0&0\\
0&0&0&0&1&1&0&0\\
0&0&0&0&0&0&1&1\\
\end{array}
\right)
\ee

\section{Unbiased diffusion}

The unbiased diffusion is defined as follows : a particle moves on the
directed De Bruijn graph $G$ and at each time step $t$, it jumps with
equal probability to one of the next neighbours of the vertex it
stands on at this time. Thus the transition probabilities matrix is
$W_0=G/2$. In the long run, the fraction of time spent on vertex $\nu$
is given by $[(W_0)^\infty]_{0,\nu}$. It can be seen (see appendix)
that 
\be
[(W_0)^k]_{\mu,\nu}=\frac{1}{2^k}\sum_{n=0}^{2^k-1}\delta_{[2^k\mu\%P]+n,\nu}.
\ee 
In particular, $(W_0^{M+k})_{\mu,\nu}=\frac{1}{P}$ for all $k\ge
0$, that is, all strings $\mu$ are visited with the same frequency
$\rho^\mu=1/P$.

In order to have a intuitive feeling of those graphs, we write them
for $M=3$ : 
\be 
W_0^2=\frac{1}{4}\left(
\begin{array}{cccccccc}
1&1&1&1&0&0&0&0\\
0&0&0&0&1&1&1&1\\
1&1&1&1&0&0&0&0\\
0&0&0&0&1&1&1&1\\
1&1&1&1&0&0&0&0\\
0&0&0&0&1&1&1&1\\
1&1&1&1&0&0&0&0\\
0&0&0&0&1&1&1&1\\
\end{array}
\right)~~~~~~
W_0^3=W_0^4=W_0^5=\ldots=\frac{1}{8}\left(
\begin{array}{cccccccc}
1&1&1&1&1&1&1\\
1&1&1&1&1&1&1\\
1&1&1&1&1&1&1\\
1&1&1&1&1&1&1\\
1&1&1&1&1&1&1\\
1&1&1&1&1&1&1\\
1&1&1&1&1&1&1\\
1&1&1&1&1&1&1\\
\end{array}
\right)
\ee 

\section{Randomly biased diffusion}

\begin{figure}
\centerline{\psfig{file=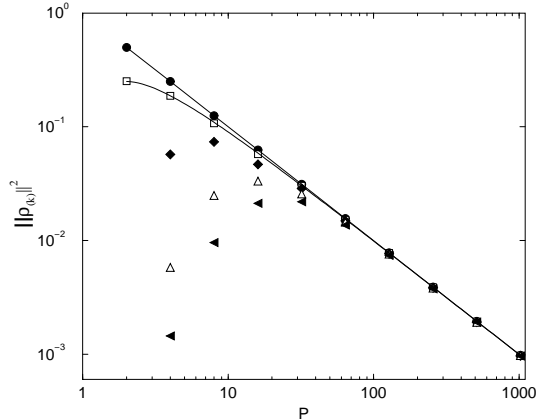,width=7cm}}
\caption{\label{rhoK}Squared norms of $\rho_{(k)}$ for k=0,\dots,4 (circles, 
squares, diamonds, triangles up, full triangles)(average over 500 samples). 
They decrease as $1/P$ for large $P$. The continuous lines are exact 
theoretical predictions.}
\end{figure}

The perturbations are introduced by adding a term to the transition
probabilities matrix $W_\eps=W_0+\eps W_1$ where $\eps$ quantifies the
asymmetry and $W_1$ contains the disorder $\xi$ 
\be
(W_1)_{\mu,\nu}=(-1)^\nu\xi_{\mu}(W_0)_{\mu,\nu} 
\ee 
where the $\xi$
are iid from the pdf $P(\xi)=1/2\ \delta(\xi-1)+1/2\ \delta(\xi+1)$
and the $(-1)^\nu$ comes from the normalization of the perturbed
probabilities. We are looking for the stationary transition
probabilities, i.e., $W_\eps^\infty$ such that
$W_\eps^\infty=\lim_{k\to\infty} \pr{W_\eps}^k$. It exists since
$W_\eps$ is a bounded operator. Its formal series expansion in $\eps$
is noted by $W_\eps^\infty=\sum_{k\ge 0}\eps^k W_{k}^\infty $ where
$W_{0}^\infty$ is a matrix whose all coefficients are equal to $1/P$
(see above). The relationship $W_\eps^\infty=W_\eps^\infty W_\eps$
provides the recurrence 
\be 
W_{k}^\infty=W_{k}^\infty
W_{0}+W_{k-1}^\infty W_1 
\ee 
Since $W_{k}^M W_{0}^\infty=0$, we
iterate $m-1$ times this equation by replacing $W_k^\infty$ with
$W_{k}^\infty W_{0}+W_{k-1}^\infty W_1$ in the r.h.s., yielding to 
\be
W_{k}^\infty=W_{k-1}^\infty W_1 V=W_{0}^\infty \cro{W_1V}^k, 
\ee 
where
$V=\sum_{c=0}^{M-1}\pr{W_0}^c$. At this point, it is useful to remark
that multiplying a matrix on the left by $W_{0}^\infty$ is equivalent
to averaging its columns : 
\be 
(W_0^\infty
A)_{\mu,\nu}=\sum_{a=0}^{P-1}
(W_{0}^\infty)_{\mu,a}A_{a,\nu}=\frac{1}{P}\sum_{a=0}^{P-1} A_{a,\nu}=
\textrm{average of the $\nu$-th column of $A$} 
\ee
thus the matrices $W_k^\infty$ consist of averages of columns of $(W_1
V)^k$. Therefore, $(W_k^\infty)_{\mu,\nu}$ is the $k$-th order
correction to the frequency of vertex $\nu$, that will be called
$\rho_{(k)}^\nu$ in the following. Note that
$\avg{\rho_{(k)}^\nu}_\xi=0$ for all $k\ge 1 $. The square root of the
second moment of $\rho_{(k)}^\nu$ averaged over the disorder gives an
indication of the typical value of $\rho_{(k)}^\nu$. In appendix
\ref{Drho} we obtain the approximation 
\be
\label{r(k)}
\avg{||\rho_{(k)}||^2}_\xi\sim\frac{(1-1/P)^k}{P}.  
\ee 
which is exact
for the first order perturbation. Therefore $\rho_{(k)}^\nu$ is of the
same order as the unperturbed $\rho_{(0)}^\nu$, thus it cannot be
neglected. Fig \ref{rhoK} shows that the behavior predicted by Eq
\req{r(k)} is indeed correct for large $P$.

Finally, one can estimate the second moment of $\rho^\nu$. If one supposes 
that the perturbations at different orders are independent, one obtains
\be\label{Dr}
\Delta\rho^2=\frac{1}{P}\sum_{\nu=0}^{P-1}\cro{\avg{[\rho^\nu]^2}_\xi-
\avg{\rho^\nu}_\xi^2}\simeq  \frac{1}{P^2}\cro{\frac{1}{1-(1-1/P)\eps^2}-1}\simeq \frac{1}{P^2}\frac{\eps^2}{1-\eps^2}
\ee

\section{Application to MG}

Let us first define the game\footnote{See refs \cite{CM99,CMZe99,CMZe99.2,CMZ99}
for more details}: MG consists of $N$ agents trying to be at 
each time step in minority. Each agent has $S$ strategies, or lookup tables $a_{i,s}$, $s=1,\cdots,S$ , and dynamically assigns
 a score to each of them
. At each time step $t$, the system's history $\mu(t)$ is made available 
to all agents; the latter use their best strategy\footnote{the one with the highest score} $s_i(t)$ 
 take the decision $a_{i,s_i(t)}^\mu(t)=+1$ or $-1$ and a market
maker sums up all decisions into the aggregate quantity
$A(t)=\sum_{i=1}^{N}a_{i,s_i(t)}^\mu(t)$ .

\begin{figure}
\centerline{\psfig{file=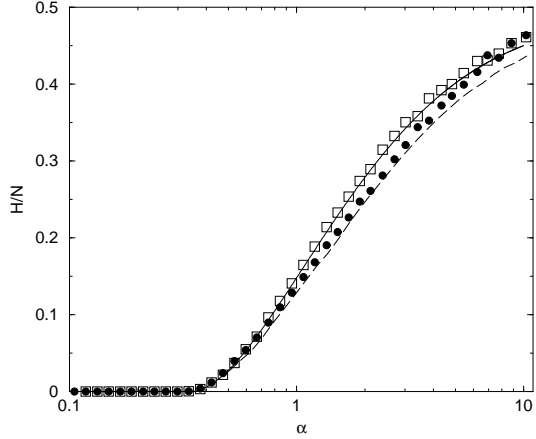,width=7cm}}
\caption{\label{H}Comparison between the available information of MG
with uniformly sampled (squares) and real histories (circles). Dashed
and continuous lines are corresponding theoretical predictions ($M=8$,
$S=2$, $300P$ iterations, average over 200 samples)}
\end{figure}

Macroscopic quantities of interest include the temporal averages of
$A(t)$ conditional to $\mu(t)=\mu$, for all $\mu$, noted by
$\avg{A^\mu}$. The MG undergoes a second order phase transition with
symmetry breaking as the control parameter $\alpha=P/N$ is varied
\cite{Savit,CM99}: the system is in the symmetric phase ($\avg{A^\mu}=0$ 
for all $\mu$) if $\alpha<\alpha_c$ and it is in the asymmetric phase
for $\alpha>\alpha_c$.
One convenient
order parameter is\footnote{$\ovl{R}=\sum_{\mu}\rho^\mu R^\mu$ is the
notation for the weighted average over the histories.}
$H=\ovl{\avg{A}^2}$: it is equal to zero in the symmetric phase, and
grows monotonically with $\alpha$ in the asymmetric phase \cite{CM99}
(see Fig. \ref{H}). One other relevant macroscopic quantity is the
fluctuations $\sigma^2=\ovl{\avg{A^2}}$ which quantifies the performance of
the agents\cite{Savit,CM99,CMZe99.2,CMZ99}.

\begin{figure}
\centerline{\psfig{file=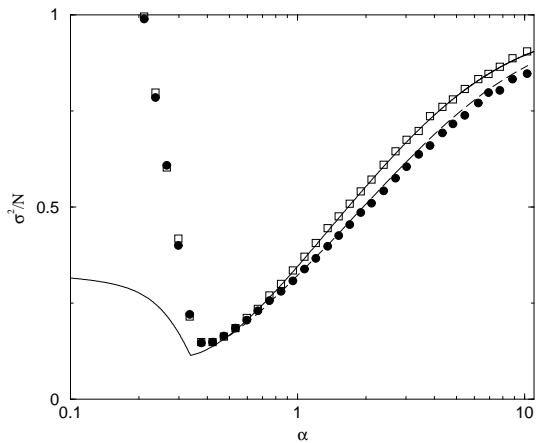,width=7cm}}
\caption{\label{s2}Comparison between the fluctuations of MG with
uniformly sampled (squares) and real histories (full circles). In the
symmetric phase, there are equal whereas they differ significantly in
the asymmetric phase. Dashed and continuous lines are corresponding
theoretical predictions; they overlap in the symmetric phase ($M=8$,
$S=2$, $300P$ iterations, average over 200 samples)}
\end{figure}

Before doing any analytic calculations, it is worth looking at Figures
\ref{s2} and \ref{H} which clearly show that Cavagna's assertion is
right as long as the system is in the symmetric phase. Indeed, if
$\avg{A^\mu}=0$, the transition probabilities from $\mu$ to its next
neighbours are unbiased, that is $\eps^\mu=0$; therefore in the
symmetric phase, where $\avg{A^\mu}=0$ for all $\mu$, the frequencies
of visit are uniform $\rho^\mu=1/P$. Accordingly, numerical
simulations show that these quantities collapse on the same curve.


\begin{figure}
\centerline{\psfig{file=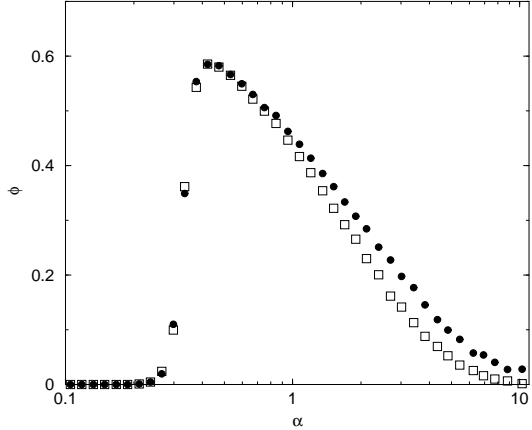,width=7cm}}
\caption{\label{phi}Comparison between the fraction of frozen agents
in MG with uniformly sampled (squares) and real histories
(circles). In the symmetric phase, there are equal whereas they differ
significantly in the asymmetric phase ($M=8$, $S=2$, $300P$
iterations, average over 200 samples)}
\end{figure}

As $\alpha$ increases, the critical point is crossed, and
$\avg{A^\mu}\ne 0$ for some $\mu$. The dynamics of the history
is biased on all such histories and consequently all macroscopic
quantities are significantly different: both $\sigma^2/N$ and $H/N$
are lower for real histories than for uniformly sampled
histories. This can be understood by the facts that $\sigma^2/N$ and
$H$ are increasing functions of $\alpha$ and that the biases on the De
Bruijn graph of histories reduce the effective number of histories,
that can be defined as $2^{-\ovl{\log_2 \rho}}$: in other words,
effective $\alpha$ of MG with real histories is smaller than that of
MG with uniform histories. This explanation is indeed confirmed by Fig
\ref{phi}; this shows the fraction of frozen agents\footnote{See
\cite{CM99}: they are agents that stop to be adaptative.} $\phi$ which
is a decreasing function of $\alpha$ in the asymmetric phase. 
As expected from the above argument, $\phi$ of MG with real histories 
is larger than that of MG with uniformly sampled histories.

The bias $\eps^\mu$ on a particular history can be estimated for large $N$: in this limit $A^\mu$ is a Gaussian variable with average $\avg{A^\mu}$ and variance $\avg{(A^\mu)^2}-\avg{A^\mu}^2$, leading to
\be\label{epsth}
\eps^\mu=\avg{\sgn(A^\mu)}\simeq\eps_{th}^\mu=\erf \pr{\sqrt{\frac{\avg{A^\mu}^2}{2[\avg{(A^\mu)^2}-\avg{A^\mu}^2]}}}.
\ee

\begin{figure}
\centerline{\psfig{file=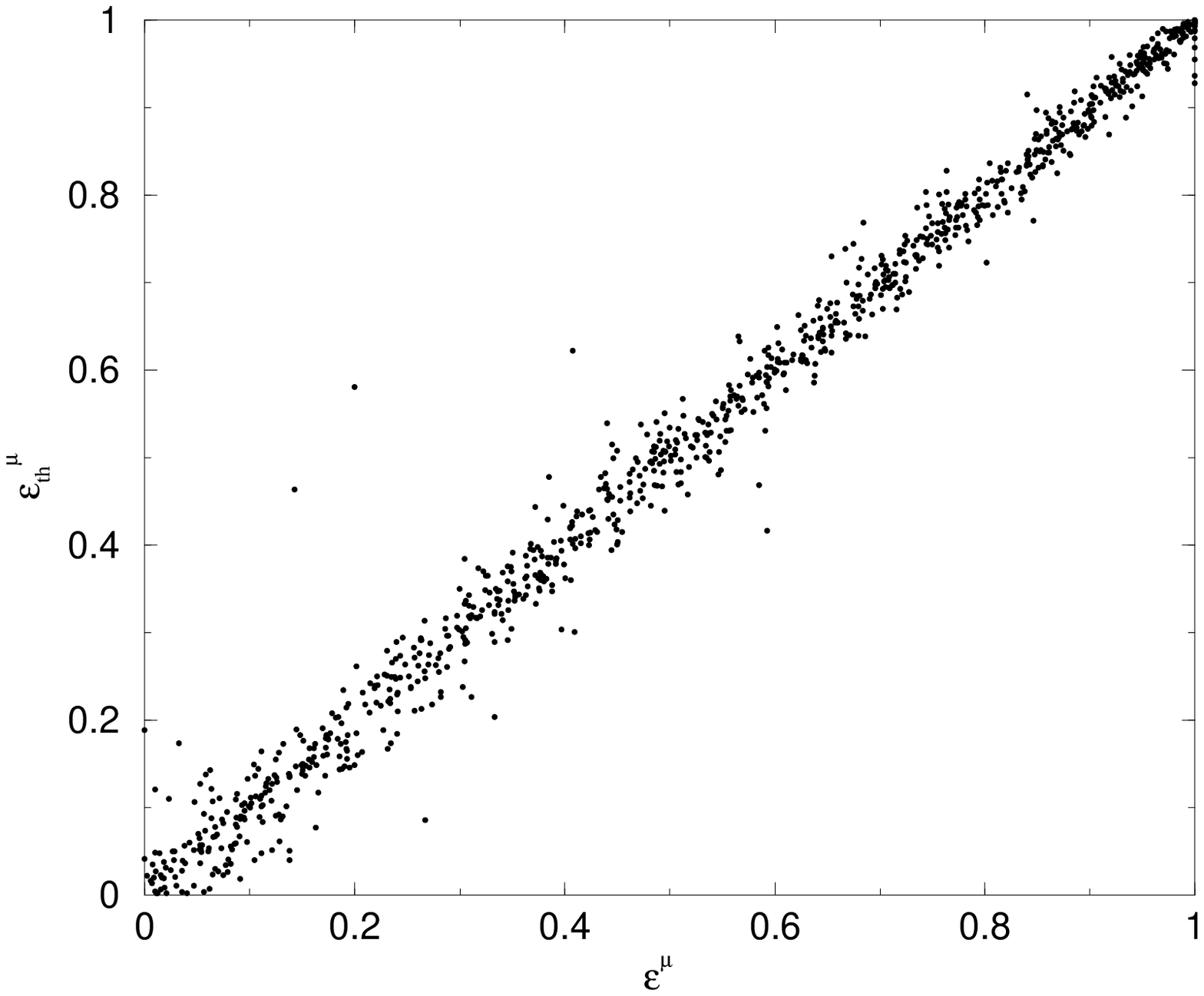,width=7cm}}
\caption{\label{eps}$\eps_{th}^\mu$ of Eq \req{epsth} vs real $\eps^\mu$ ($M=10$, $N=121$, $S=2$, $1000P$ iterations)}
\end{figure}

\begin{figure}
\centerline{\psfig{file=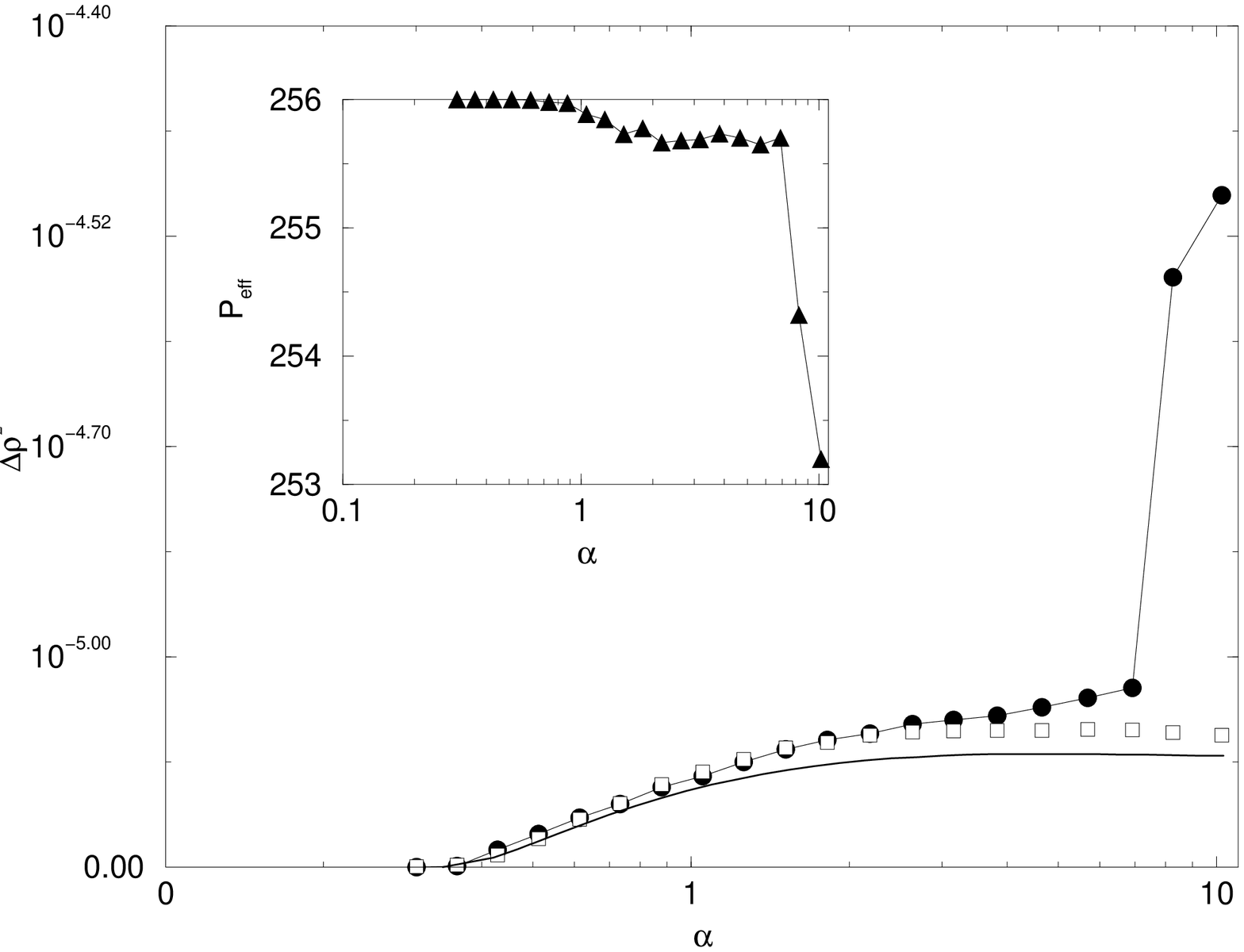,width=7cm}}
\caption{\label{Deltarho}Inhomogeneity of the frequency of histories
$\Delta\rho^2$ versus $\alpha=P/N$ from: numerical simulations (full
circles), Eq \req{Dr} with $\eps$ from numerical simulations (void
squares) and Eq \req{Dr} with $\beps$ (continuous line); inset:
average number of visited histories versus $\alpha$; ($M=8$, $S=2$,
300$P$ iterations, average over 200 samples).}
\end{figure}

Fig \ref{eps} confirms the validity of Eq. \req{epsth}. The figure also 
shows that $\eps^\mu$ are unevenly distributed and they 
are not equal even if the system is deep in the asymmetric phase
($\alpha\simeq8.5$ in this figure). Indeed, as a function of $\mu$,
$\avg{A^\mu}$ is a random variable with average 0 and variance $H$,
which is an increasing function of $\alpha$. Since we studied
diffusion of perturbed graphs with only one parameter $\eps$, we have
to map all $\eps^\mu$ onto a scalar quantity, so that we define
$\eps$ as the non weighted average\footnote{This is clearly an important
assumption, but the diffusion on De Bruijn graphs with one $\eps^\mu$
per site leads to a much greater complexity. As it appears on Fig
\ref{s2}, \ref{H} and \ref{Deltarho}, this assumption is not
unrealistic.} of $\eps^\mu$ over the histories. For large $P$, $\eps$
can be approximated by

\be\label{beps}
\beps=2\int_0^\infty\dd A\ \frac{e^{-\frac{A^2}{2H}}}{\sqrt{2\pi H}}\ 
\erf\pr{\frac{A}{\sqrt{2(\sigma^2-H)}}}.
\ee
Here both $H$ and $\sigma^2$ can be computed analytically 
with the method of refs. \cite{CMZe99,CMZ99,CMZe99.2} (see the
appendix). 
However the solution depends on the distribution $\rho^\mu$.
In order to make Eq. \req{beps} a self-consistent equation
for $\beps$, we need to parameterize the distribution of 
$\rho^\mu$ by $\beps$ itself. 

\begin{figure}
\centerline{\psfig{file=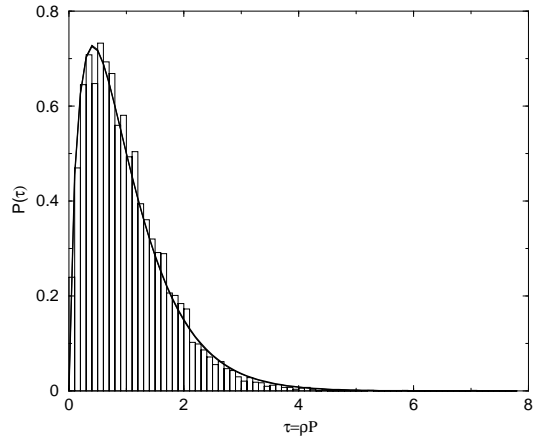,width=7cm}}
\caption{\label{distrrho}Distribution of the frequency of visit of the
histories in the minority game. The continuous line is the best fit for a pdf given by Eq \req{tau}  ($M=13$,
$N=801$, $S=2$, $400P$ iterations)}
\end{figure}

We could not find {\em ab initio} the analytic form of the pdf of 
$\{\rho^\mu\}$, but Fig \ref{distrrho} shows that 

\be\label{tau}
P(\tau)\simeq\frac{(\lambda+1)^{\lambda+1}}{\Gamma(\lambda+1)}
\tau^{\lambda} e^{-(\lambda+1)\tau}
\ee
is a very good approximation for the pdf of $\rho=\tau/P$. 
The parameter $\lambda$ is easily connected with $\beps$: 
\be
\avg{\tau^2}-\avg{\tau}^2=\frac{1}{1+\lambda}=P^2\Delta\rho^2\simeq\frac{\beps^2}{1-\beps^2}
\ee
where we used Eq. (10). This gives $\lambda\simeq(1-2\beps^2)/\beps^2$. Note
that this approximation requires $\beps<1/\sqrt{2}$.

\begin{figure}
\centerline{\psfig{file=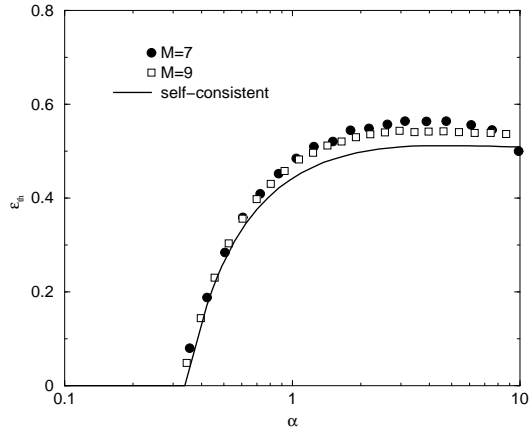,width=7cm}}
\caption{\label{erf}$\eps$ versus $\alpha=P/N$ ($M=8$, $S=2$, $300P$
iterations, average over 200 samples). The straight line is $\beps$,
the theoretical prediction of the self consistent theory.}
\end{figure}

This turns Eq. \req{beps} into an equation for $\beps$, and 
the theory is self consistent. Figure \ref{erf}
reports measured $\eps$ and its approximation $\beps$. What clearly
appears from this figure is that $\eps$ is far from being
negligible, and that $\beps$ is a quite good approximation to
$\eps$.

We can also check the validity of Eq. \req{Dr} against the
self-consistent theory. Fig \ref{Deltarho} shows
that Eq. \req{Dr} is in good agreement with numerical simulations as
long as all histories are visited. Moreover the approximation $\beps$
for $\eps$ leads to qualitatively similar results, but underestimates
$\Delta \rho^2$ because $\beps<\eps$ (see figure \ref{erf}).

The self-consistent replica calculation for the Minority Game
of refs. \cite{CMZe99,CMZ99,CMZe99.2} with the ansatz
$\rho=\tau/P$ and $\tau$ given by the pdf \req{tau} is discussed in the
appendix. Fig \ref{s2} and
\ref{H} indicate that analytic predictions are well supported by
numerical simulations.

In the asymmetric phase, which is arguably the most relevant and interesting 
in the MG \cite{CMZe99.2}, all quantities of MG change significantly if one
replaces real histories with random uniform histories. 
A dependence on the frequencies $\rho^\mu$ does not necessarily
imply the relevance of the detailed dynamics of the histories. 
If the histories $\mu$ where drawn randomly from the ``correct'' 
distribution $\rho^\mu$, the results would be the same (actually it 
suffices to know the pdf of $\rho^\mu$).
The problem is that 
the distribution $\rho^\mu$ depends on the asymmetry $\avg{A^\mu}$,
which in turn depend on the microscopic constitution of all agents
\cite{CM99}. In other words, $\rho^\mu$ 
is a self-consistently determined quantity
and hence it is only known {\em a posteriori}.

\section{Conclusion}

We have shown that the dynamics of histories cannot be considered as irrelevant. Indeed, even for the canonical MG, it is relevant and cannot be replaced by randomly drawn histories. In addition, for many extensions and variations of the MG, the dynamics of histories is not only relevant, but crucial.

We acknowledge fruitful discussions with Philippe Flajolet and Paolo De Los Rios.
This work has been partially supported by the Swiss National Science Foundation under Grant Nr 20-46918.98.

\appendix

\section{}\label{Wn}

Let us prove by induction that
\be
(W_0^k)_{\mu,\nu}=\frac{1}{2^k}\sum_{n=0}^{2^k-1}\delta_{[2^k\mu\%P]+n,\nu}
\ee

It is sufficient to calculate explicitly $(W_0^k)_{\mu,\nu}$ from $(W_0^{k-1})_{\mu,\nu}$
\bea
(W_0^k)_{\mu,\nu}&=&\sum_{\tau=0}^{P-1}(W_0)_{\mu,\tau}(W_0^{k-1})_{\tau,\nu}\nonumber\\
&=&\sum_{n=0}^{2^{k-1}-1}\acc{\delta_{[2^{k-1}([2\mu\%P])\%P]+n,\nu}+\delta_{[2^{k-1}([2\mu\%P]+1)\%P]+n,\nu}}\nonumber\\
&=&\frac{1}{2^k}\sum_{n=0}^{2^k-1}\delta_{[2^k\mu\%P]+n,\nu}
\eea

since  $A(B\%P)\%P=AB\%P$ and  $(2^k\mu+2^{k-1})\%P=[2^k\mu\%P]+2^{k-1}$ if $P=2^M$ and  $k\le m-1$. 

\section{}\label{Drho}

In order to simplify the notations, we define
\be
(X^c)_{\mu,\nu}=\sum_{n=0}^{2^c-1}\delta_{[2^{c+1}\mu\%P]+n,\nu}-\delta_{[2^{c+1}\mu\%P]+n+2^c,\nu}
\ee
This matrix is such that
\be
(X^c)_{\mu,\nu}=\left\{\begin{array}{rcl}1 &\textrm{if} &2^{c+1}\mu\%P\le\nu<[2^{c+1}\mu\%P]+2^c\}\\-1&\textrm{if}&[2^{c+1}\mu\%P]+2^c\le\nu<2^{c+1}(\mu+1)\%P\\0&\textrm{else}&\end{array}\right.
\ee
With this formalism, one can write $W_1 V$ as 
\be\label{r1nu}
(W_1V)_{\mu,\nu}=\frac{\xi_\mu}{2}\sum_{c=0}^{M-1}\frac{1}{2^c}(X^c)_{\mu,\nu}
\ee

Let us calculate the perturbation at order 1: one has to compute $||\rho_{(1)}||^2$ in order to have an estimation of  the typical value of a generic $\rho_{(1)}^\nu$: since the $\xi$ are uncorrelated and $\sum_{\nu=0}^{P-1}(X^c)_{\mu,\nu}(X^d)_{\mu,\nu}=2^{c+1}\delta_{c,d}$ ,
\be
\avg{||\rho_{(1)}||^2}_\xi=\frac{1}{4 P^2}\sum_{\mu,\nu=0}^{P-1}\sum_{c=0}^{M-1}\frac{[(X^c)_{\mu,\nu}]^2}{2^{2c}}=\frac{(1-1/P)}{P}
\ee

The next orders of perturbation are much harder to handle. However, for large $P$, one can  approximate them by supposing that 
\be
\avg{||\rho_{(k)}||^2}_\xi\sim (1-1/P) \avg{||\rho_{(k-1)}||^2}_\xi=\frac{(1-1/P)^k}{P}.
\ee
 Consequently, $\rho_{(k)}^\nu\sim (1-1/P)^{k/2}\frac{1}{P}\simeq\frac{1}{P}$ at leading order 

\section{}\label{repl}

Since agents actually minimize $H/N$, one can consider this quantity as a Hamiltonian and find its ground state. This is possible by methods of statistical physics such as replica trick \cite{MPV,dotsenko}.
The generalization of the calculus of refs \cite{CMZe99,CMZ99,CMZe99.2} to $\rho^\mu=\tau^\mu/P$ drawn from the pdf given by Eq \req{tau} and \req{beps} is straightforward; the free energy reads in the thermodynamic limit
\bea
F(\beta,Q,q,R,r)&=&\Avg{\frac{\alpha}{2\beta}\log[1+\chi\tau]}_\tau+\frac{1+q}{2}\Avg{\frac{1}{\frac{1}{\tau}+\chi}}_\tau\nonumber\\&+&\frac{\alpha\beta}{2}(RQ-rq)-\frac{1}{\beta}
\avg{\log\int_{-1}^1 ds e^{-\beta  (\zeta s^2-\sqrt{\alpha r}\,z\,s)}}_z
\label{FRS}
\eea
where  $\chi=\beta(Q-q)/\alpha$ and $\zeta=-\sqrt{\alpha/r}\ \beta(R-r)$. Next, the $\beta\to\infty$ limit is taken while keeping finite $\chi$ and $\zeta$. One obtains
\be\label{Hth}
H=\frac{1+Q}{2}\cro{\Avg{\frac{1}{\frac{1}{\tau}+\chi}}_\tau-\chi\Avg{\frac{1}{[\frac{1}{\tau}+\chi]^2}}_\tau}
\ee
and 
\be\label{s2th}
\sigma^2=H+\frac{1-Q}{2}
\ee
where $Q$ and $\chi$ take their saddle point values, given by the solution of

\bea
\label{Q} Q(\zeta)&=&1-\sqrt{\frac{2}{\pi}}\frac{e^{-\zeta^2/2}}{\zeta}-\pr{1-\frac{1}{\zeta^2}}\erf\pr{\frac{\zeta}{\sqrt{2}}}\\
\label{zeta}
Q(\zeta)&=&\frac{1}{\alpha}\cro{\frac{\erf (\zeta/\sqrt{2})}{\chi\zeta}}^2\ \frac{1}{\avg{\frac{1}{[1/\tau+\chi]^2}}_\tau}-1\\
\label{chizeta}\chi \Avg{\frac{1}{\frac{1}{\tau}+\chi}}_\tau&=&\frac{\erf(\zeta/\sqrt{2})}{\alpha}
\eea
Eqs \req{zeta} and \req{chizeta}, together with Eq \req{beps}, form a closed set of equations that has to be solved numerically. Note that as in the random histories case, $\chi$ becomes infinite at the critical point, where $\alpha_c=\erf(\zeta/\sqrt{2})$.

\end{document}